\documentclass[11pt,a4paper]{article}

\pdfoutput=1

\usepackage{jcappub}
\usepackage{hyperref}
\usepackage{amssymb,esint}
\usepackage{graphicx}
\usepackage{epsfig}

\title{The integrated Sachs-Wolfe imprint of cosmic superstructures: a problem for $\Lambda$CDM}

\author[a,b]{Seshadri Nadathur,}
\author[c]{Shaun Hotchkiss}
\author[a]{and Subir Sarkar}

\affiliation[a]{Rudolf Peierls Centre for Theoretical Physics, University of
Oxford, Oxford OX1 3NP, UK}
\affiliation[b]{Fakult\"at f\"ur Physik, Universit\"at Bielefeld, Postfach 100131, 33501 Bielefeld, Germany}
\affiliation[c]{Department of Physics, University of Helsinki and Helsinki
Institute of Physics, P.O. Box 64, FIN-00014 University of Helsinki, Finland}

\emailAdd{seshadri@physik.uni-bielefeld.de}
\emailAdd{shaun.hotchkiss@helsinki.fi}
\emailAdd{s.sarkar@physics.ox.ac.uk}

\abstract{A crucial diagnostic of the $\Lambda$CDM cosmological model is the
integrated Sachs-Wolfe (ISW) effect of large-scale structure on the
cosmic microwave background (CMB). The ISW imprint of superstructures
of size $\sim100\;h^{-1}$Mpc at redshift $z\sim0.5$ has been detected
with $>4\sigma$ significance, however it has been noted that the
signal is much larger than expected. We revisit the calculation using
linear theory predictions in $\Lambda$CDM cosmology for the number
density of superstructures and their radial density profile, and take 
possible selection effects into account. While our expected signal is
larger than previous estimates, it is still inconsistent by $>3\sigma$
with the observation. If the observed signal is indeed due to
the ISW effect then huge, extremely underdense voids are far more common in
the observed universe than predicted by $\Lambda$CDM.}


\arxivnumber{1109.4126}

\begin{document}
\maketitle
\flushbottom


\section{Introduction}
\label{section:intro}

The standard `concordance' $\Lambda$CDM cosmological model fits many
different observations, including the luminosity distance-redshift
relation of Type Ia supernovae (SNe~Ia) 
(e.g.~\cite{Hicken:2009dk,Amanullah:2010vv,Conley:2011ku}), the
anisotropies in the CMB \cite{Komatsu:2010fb}, the locally measured
Hubble parameter (e.g.~\cite{Riess:2011yx}) and baryon acoustic
oscillations (e.g.~\cite{Percival:2009xn} --- but 
see~\cite{Labini:2009ke,Kazin:2009cj}). These observations, when 
interpreted assuming the homogeneous and isotropic 
Friedman-Robertson-Walker metric, imply that the expansion of the
universe is accelerating, from which it is inferred that the universe
is presently dominated by a cosmological constant (`dark energy') with
negative pressure. It is important to note that this evidence is
mostly \emph{geometrical}, being based on interpreting measurements of
distances --- made using `standard rulers' (the sound horizon at last
scattering) and `standard candles' (SNe~Ia) --- as due to accelerated
expansion.~The same data can equally be fitted without dark
energy if, e.g., the isotropic but radially inhomogeneous
Lema\"itre-Tolman-Bondi metric is assumed and other assumptions such
as a power-law spectrum for the primordial density perturbations are
relaxed (e.g.~\cite{Biswas:2010xm,Nadathur:2010zm}).

Given that dark energy is a complete mystery from a physical
viewpoint, it is therefore imperative to examine the observational
evidence for its \emph{dynamical} effects. For example, the decay of
gravitational potentials after dark energy begins to dominate (at
redshift $z \lesssim 1$) should lead to secondary CMB anisotropies as
the CMB photons traverse regions of over- or under-density --- the ISW
effect \cite{Sachs:1967er}. If the universe is spatially flat, then 
detection of the ISW effect through cross-correlation of the CMB with 
large-scale structure would provide direct evidence of dark energy's 
negative pressure, hence crucial confirmation of the $\Lambda$CDM 
model \cite{Crittenden:1995ak}.\footnote{The ISW effect should also boost low
multipoles in the CMB angular power spectrum, whereas these are in fact
anomalously low on the observed sky. However given the large cosmic variance
on these scales and galactic foreground systematics, this discrepancy 
is not thought to be significant~\cite{Bennett:2010jb}.}

To detect the ISW effect with 5$\sigma$ significance in CMB-galaxy
cross-correlations requires $z$ measurements for over 10 million
galaxies \cite{Afshordi:2004kz,Douspis:2008xv}. Such datasets are not
yet available but several authors
(e.g.,~\cite{Fosalba:2003ge,Boughn:2003yz,Afshordi:2003xu,Nolta:2003uy,
Padmanabhan:2004fy,Giannantonio:2006du,Cabre:2006qm,Raccanelli:2008bc})
have examined smaller source catalogues and reported
detections with $<3\sigma$ significance; however, \cite{HernandezMonteagudo:2009fb} 
provides a skeptical view of some of these analyses. 
Conversely, some authors \cite{Rassat:2006kq,Francis:2009ps} were 
unable to reject the null hypothesis (no ISW effect) and others 
\cite{Sawangwit:2009gd} even found a slight \emph{anti}-correlation 
rejecting $\Lambda$CDM at $2-3\sigma$ significance. Some 
groups have combined different data sets to increase the detection 
significance above $4\sigma$ \cite{Ho:2008,Giannantonio:2008zi} but it
has been argued that these analyses have underestimated the error bars 
\cite{LopezCorredoira:2010rr}.

Much of the uncertainty in full-sky studies arises from the difficulty
in reconstructing the underlying density field from galaxy survey
data, given Poisson noise in the galaxy distribution. A different
approach to this problem is followed by Granett \emph{et al.} 
\cite{Granett:2008ju,Granett:2008xb}, who study the Sloan Digital Sky
Survey (SDSS) Data Release 6 (DR6) luminous red galaxies (LRGs).~They
use 3D galaxy information rather than the projected 2D density, and
select only the most extreme density perturbations, which are
unambiguously identified despite Poisson noise. Along the lines of
sight corresponding to these `superstructures' they report a
$4.4\sigma$ detection of the ISW effect, which is the most
significant reported detection to date. This approach also provides information
about the sizes and distribution of extreme structures in the universe
so can be used to check the consistency of the standard $\Lambda$CDM
model of structure formation, in particular whether the primordial
fluctuations were indeed gaussian.

However the magnitude of the temperature signal reported in
Ref.~\cite{Granett:2008ju} is surprisingly large and has
been argued \cite{Hunt:2008wp,Inoue:2010rp} to be quite inconsistent
with $\Lambda$CDM. In response, it has been noted \cite{Papai:2010eu} that
the assumed profile of the superstructures has a big effect on the
signal, and that the `compensated top-hat' profile adopted in 
Refs.~\cite{Hunt:2008wp,Inoue:2010rp} is not the most appropriate. Using an 
alternative profile that is calibrated against $N$-body simulations 
and adopting a template-fit approach, it is claimed \cite{Papai:2010gd} 
that the discrepancy with $\Lambda$CDM is only at the $2\sigma$ level.

Our aim in this paper is to clarify this important issue. We calculate
the expected temperature signal from these superstructures making no
\emph{a priori} assumptions about their nature except that they arose
in a $\Lambda$CDM cosmology with gaussian primordial density
perturbations. Our analysis uses linear perturbation theory, but non-linear
corrections are argued to be sub-dominant at the relevant length scales 
and redshift. We use a statistical treatment of (initially) gaussian
perturbations \cite{Bardeen:1985tr} to calculate the expected number densities
and the expected profiles of the superstructures. These profiles are not 
significantly different from those used in Ref.~\cite{Papai:2010eu}, and we do find 
a factor of $\sim3$ increase in the ISW signal over that calculated earlier 
\cite{Hunt:2008wp,Inoue:2010rp}. However, contrary to Ref.~\cite{Papai:2010gd} 
we find that the expected value of the signal is \emph{still} discrepant at 
$>3\sigma$ with the observations reported by G08a. We demonstrate 
that this difference in the conclusions arises due to the 
interpretation of the template-fitting approach used in 
Ref.~\cite{Papai:2010gd} and argue that a correct interpretation would lead 
to conclusions compatible with those presented here.

In Section~\ref{section:ISW} we briefly review the ISW effect and in
Section~\ref{section:LCDMprediction} we calculate the expected
temperature signal of superstructures in the standard $\Lambda$CDM
model. In Section~\ref{section:observation} we describe the key
features of the observational strategy \cite{Granett:2008ju,Granett:2008xb} 
which must be accounted for before making a comparison with the 
theoretical calculation. In Section~\ref{section:comparison} we show that 
even if Refs.~\cite{Granett:2008ju,Granett:2008xb} had
selectively picked out the regions in the survey with the biggest ISW
signal, there is still a significant discrepancy with the $\Lambda$CDM 
expectation. In Section~\ref{subsection:differences} we discuss the reasons 
for the difference between our result and that of Ref.~\cite{Papai:2010gd}. 

Finally in Section~\ref{section:conclusions} we turn to possible alternative 
explanations outside the standard cosmological model. Modifications to the 
growth rate of perturbations from the $\Lambda$CDM prediction could be 
responsible for the discrepancy, but any alternative model must also match the 
other observational successes of $\Lambda$CDM. A plausible explanation 
is that the effect is due to non-gaussianity of the primordial perturbations, but 
further work is needed to calculate its effect on the expected signal. Until this 
discrepancy is clarified, the observations reported in Ref.~\cite{Granett:2008ju} 
do not appear to be compatible with the ISW effect expected in the standard 
cosmology.


\section{The ISW effect}
\label{section:ISW}

In a universe with matter density $\Omega_\mathrm{m}=1$ and no dark
energy, density perturbations $\delta$ grow at exactly the same rate
as the scale factor of the universe $a~(\equiv (1+z)^{-1})$, so at the
linear level there is no evolution of the gravitational potential
$\Phi~(\propto -\delta/a)$. However, in a $\Lambda$CDM universe, $a$
grows faster than (linear) density perturbations, so perturbations in
$\Phi$ decay with time. For a CMB photon passing through an overdense
region the energy gained while falling in is not cancelled by the
energy lost in climbing out of the evolved, shallower, potential
well. Overdense regions (clusters) therefore appear as hot spots in
the CMB; conversely, underdense regions (voids) will appear as cold
spots as the photon loses more energy climbing the potential hill than
it gains subsequently while descending.

The temperature fluctuation $\Delta T(\hat{n})$ induced along
direction $\hat{n}$ is \cite{Sachs:1967er}:
\begin{equation}
\label{eq:SWintegral}
\Delta T(\hat{n}) = \frac{2}{c^3} \bar{T}_0 \int_{0}^{r_\mathrm{L}} \dot 
 \Phi(r, z, \hat{n})\;a\;\mathrm{d}r\,,
\end{equation}
where $\bar{T}_0$ is the mean CMB temperature, $r_\mathrm{L}$ is the
radial comoving distance to the last scattering surface (LSS),
$\dot\Phi(r,z,\hat{n})$ is the time derivative of the gravitational
potential along the photon geodesic and $c$ is the speed of light.

The Poisson equation relates $\Phi$ to the density contrast
$\delta\equiv(\rho-\bar{\rho})/\bar{\rho}$ (where $\bar{\rho}(t)$ is
the mean density) through:
\begin{equation}
\label{eq:Poisson}
\nabla^2\Phi({\bf x},t) = 4\pi G\bar{\rho}(t) a^2 \delta({\bf x}, t).
\end{equation}
This can be written in Fourier space as
\begin{equation}
\label{eq:FourierPoisson}
\Phi({\bf k},t) = -\frac{3}{2}\left(\frac{H_0}{k}\right)^2 \Omega_\mathrm{m}
 \frac{\delta({\bf k}, t)}{a}\,,
\end{equation}
where $H_0$ is the current Hubble parameter. Taking the time
derivative of this equation yields:
\begin{equation}
\label{eq:timederiv}
\dot\Phi({\bf k}, t) = \frac{3}{2}\left(\frac{H_0}{k}\right)^2\Omega_\mathrm{m} 
 \left[\frac{\dot{a}}{a^2}\delta({\bf k}, t) 
 - \frac{\dot\delta({\bf k}, t)}{a}\right].
\end{equation}

We assume that linear theory holds on the large scales of interest
hence perturbations grow as $\delta({\bf k},t)=D(t)\delta({\bf
  k},z=0)$, where $D(t)$ is the linear growth factor. A numerical
simulation has shown that non-linear effects represent only a $10\%$
correction at the low redshifts we are interested
\cite{Cai:2010hx}.\footnote{Both the linear and non-linear effects
  grow with time, however at late times and large scales ($\sim100
  \;h^{-1}\mathrm{Mpc}$) the linear effect dominates while at early times
  (when $\Omega_\Lambda\simeq0$), both effects are smaller but the
  non-linear effect dominates.} In this approximation,
\begin{equation}
\label{eq:lineartimederiv}
\dot\Phi({\bf k},z) =
\frac{3}{2}\left(\frac{H_0}{k}\right)^2\Omega_\mathrm{m}\frac{H(z)}{a}
\left[1 - \beta(z)\right]\delta({\bf k}, z)\,,
\end{equation}
where $\beta(z)\equiv \mathrm{d}\ln D/\mathrm{d}\ln a$ is the linear
growth rate. Hence the time evolution is captured by the ISW linear
growth factor, $G(z) = H(z)\left(1 - \beta(z)\right)D(z)/a$. For an
$\Omega_\mathrm{m} = 1$ universe, $\beta(z) = 1$ for all $z$ so there
is no ISW effect.

Given the density profile $\delta$ of any isolated superstructure,
Eqs.~(\ref{eq:SWintegral}) and (\ref{eq:lineartimederiv}) can be used
to calculate the temperature fluctuation it induces in the
CMB. Assuming spherical symmetry of the density profile,
Eq.~(\ref{eq:lineartimederiv}) in real-space becomes:
\begin{equation}
\label{eq:realspace}
\dot\Phi(r, z) = \frac{3}{2}\Omega_\mathrm{m} H_0^2 G(z) F(r)\,,
\end{equation}
where
\begin{equation}
\label{eq:F(r)}
F(r) = \int_0^r \frac{r^{\prime2}}{r}\delta(r^\prime)\;\mathrm{d}r^\prime + 
 \int_r^\infty r^{\prime}\delta(r^\prime)\;\mathrm{d}r^\prime\,,
\end{equation}
with $\delta(r^\prime)$ evaluated at redshift $z=0$. Thus $F(r)$
contains all information about the structure in question, while the
assumed cosmology enters through the prefactor and the ISW growth
factor $G(z)$ in Eq.~(\ref{eq:realspace}).


\section{Expected signal from superstructures in $\Lambda$CDM}
\label{section:LCDMprediction}

`Superstructures' refer to density perturbations extending over
$\gtrsim100\;h^{-1}$Mpc and should not be thought of as non-linear
collapsed structures in the usual sense, rather as smooth hills and
valleys in the density distribution. Collapsed structures form only
where the density perturbation $\delta({\bf r)}$ exceeds unity, which
happens on much smaller scales than those of interest here.

It is stated in Ref.~\cite{Granett:2008ju} that the most extreme structure in the
$(500\;h^{-1}\mathrm{Mpc})^3$ box of the Millennium $N$-body
simulation~\cite{Springel:2005nw}, when placed at $z=0$, would
produce a signal of $\Delta T\sim4.2\;\mu$K. However when
the signal distribution for supervoids with the densities and sizes reported in Ref.~\cite{Granett:2008xb} was calculated, the answer was only
$\langle\Delta T\rangle =-0.42\;\mu$K \cite{Hunt:2008wp}. These 
authors \cite{Hunt:2008wp} assumed a `compensated
top-hat'  density profile motivated by the
asymptotic final state of a void \cite{Sheth:2003py}. A similar
profile was assumed by other authors \cite{Inoue:2010rp} who found a similar
average signal $\langle\Delta T\rangle=-0.51\;\mu$K for the 50 most
extreme density perturbations of fixed radius $r=130\;h^{-1}$Mpc
expected in a $\Lambda$CDM cosmology. Subsequently it was argued 
\cite{Papai:2010eu,Papai:2010gd} that this profile
is not the appropriate choice for density perturbations on
$\gtrsim100\;h^{-1}$Mpc scales; instead an \emph{uncompensated} density 
profile was motivated from gaussian statistics, and was found \cite{Papai:2010eu}
to give larger values of $\Delta T$. Using this profile Ref.~\cite{Papai:2010gd} 
claimed only a $2\sigma$ discrepancy between the $\Lambda$CDM
prediction and observation. However, as we discuss 
in Section~\ref{subsection:differences}, the interpretation of \cite{Papai:2010gd} 
implies underdense regions with a physical density contrast 
$\delta<-1$, which is physically impossible.

It is thus necessary to revisit this issue. Using the statistics of a
homogeneous, isotropic, gaussian density field, we now derive the
expected mean density profiles of superstructures of all density
contrasts and all sizes, as well as the expected number density of
such superstructures.

\subsection{The number density of structures on different scales}
\label{subsection:numberdensity}

We identify superstructures of different sizes with extrema of the
linear density perturbation field $\delta({\bf r)}$ when smoothed over
different scales. Overdensities correspond to peaks of the smoothed
field and underdensities to troughs. In the $\Lambda$CDM model,
$\delta({\bf r})$ is a homogeneous and isotropic, gaussian-distributed
random field and the statistical properties of the maxima and minima
have been calculated in Ref.~\cite{Bardeen:1985tr}. We
briefly review below their key results and introduce necessary
notation.

Let $P(k,t)$ denote the matter power spectrum, defined as the Fourier
transform of the 2-point correlation function $\xi(r,t)$ of the
density field at time $t$. Define a set of spectral moments weighted
by powers of $k$:
\begin{equation}
\label{eq:sigmaj}
\sigma_j^2(t) = \int \frac{k^2\;\mathrm{d}k}{2\pi^2} W^2 (kR_\mathrm{f}) 
 P(k, t) k^{2j}\,,
\end{equation}  
where $W(kR_\mathrm{f})$ is the window function appropriate to the
filter used to smooth the density field, and $R_\mathrm{f}$ is the
(comoving) smoothing scale. Thus $\sigma_0$ is just the standard rms
fluctuation of the smoothed density field. Using a gaussian filter,
$W(kR_\mathrm{f}) = \exp(-k^2R_\mathrm{f}^2/2)$, we define the
parameters:
\begin{equation}
\label{eq:gammaRstar}
\gamma\equiv\frac{\sigma_1^2}{\sigma_2\sigma_0}\,,\,\,\,
 R_\ast\equiv\sqrt{3}\frac{\sigma_1}{\sigma_2}\;.
\end{equation}
The (comoving) differential number density
$\mathcal{N}_\mathrm{max}(\nu)$ of maxima of height
$\delta_0=\nu\sigma_0$ is then \cite{Bardeen:1985tr}:
\begin{equation}
\label{eq:diffnumberdensity}
\mathcal{N}_\mathrm{max}(\nu)\mathrm{d}\nu = \frac{1}{(2\pi)^2 R_\ast^3} 
 \mathrm{e}^{-\nu^2/2}G(\gamma,\gamma\nu)\mathrm{d}\nu\;.
\end{equation}
The function $G(\gamma,\gamma\nu)$ is given by Eq. (A19) of 
Ref.~\cite{Bardeen:1985tr}; we use a fitting form, accurate to better than 
1\%, given in their Eqs.~(4.4) and (4.5). The density of minima is related simply to that
of maxima through:
$\mathcal{N}_\mathrm{min}(\nu)=\mathcal{N}_\mathrm{max}(-\nu)$.

\subsection{Mean radial profiles}
\label{subsection:radialprofiles}

Having identified superstructures with the maxima or minima in the
smoothed density field, we wish to determine the mean radial variation
of the density field in the neighbourhood of these extrema. Ref.~\cite{Bardeen:1985tr}
shows that, given a maximum $\delta=\delta_0$ at ${\bf r}=0$, the mean shape
in the vicinity of this point after averaging over all possible
orientations of the principal axes as well as all values of the
curvature at ${\bf r}=0$ is:
\begin{equation}
\label{eq:BBKS7.10}
\bar\delta(r)=\frac{\delta_0}{(1 - \gamma^2)} \left(\psi+\frac{R_\ast^2}{3}
 \nabla^2\psi\right) - \frac{\langle x|\delta_0\rangle\sigma_0}{\gamma(1 - 
 \gamma^2)}\left(\gamma^2\psi + \frac{R_\ast^2}{3}\nabla^2\psi\right)\;,
\end{equation}
where $\psi(r)\equiv\xi(r)/\xi(0)$ is the normalised density-density
correlation function, and $x=-\nabla^2\delta/\sigma_2$. The expectation
value of $x$ given a peak of height $\delta_0$ at ${\bf r}=0$ is
approximately
\begin{equation}
\label{eq:x}
\langle x|\delta_0\rangle = \gamma\nu + \Theta(\gamma,\gamma\nu)\;,
\end{equation}
where $\nu=\delta_0/\sigma_0$, and $\Theta(\gamma,\gamma\nu)$ is given
by the fitting function (6.14) of Ref.~\cite{Bardeen:1985tr}. It follows that
Eq.~(\ref{eq:BBKS7.10}) can be rewritten \cite{Lahav:1991} as:
\begin{equation}
\label{eq:LiljeLahavform}
\bar\delta(r,t)=\frac{1}{\sigma_0}\int_0^\infty
 \frac{k^2}{2\pi^2}\frac{\sin(kr)}{kr}W^2(kR_\mathrm{f})P(k,t)
 \left[\frac{\nu-\gamma^2\nu-\gamma\Theta}{1-\gamma^2} + 
 \frac{\Theta R_\ast^2k^2}{3\gamma(1-\gamma^2)}\right]\mathrm{d}k,
\end{equation}
which we use for numerical evaluation of profiles. Note that $\gamma$
and $R_\ast$ depend on the smoothing scale $R_\mathrm{f}$.

\begin{figure}
\center
\includegraphics[width=\columnwidth]{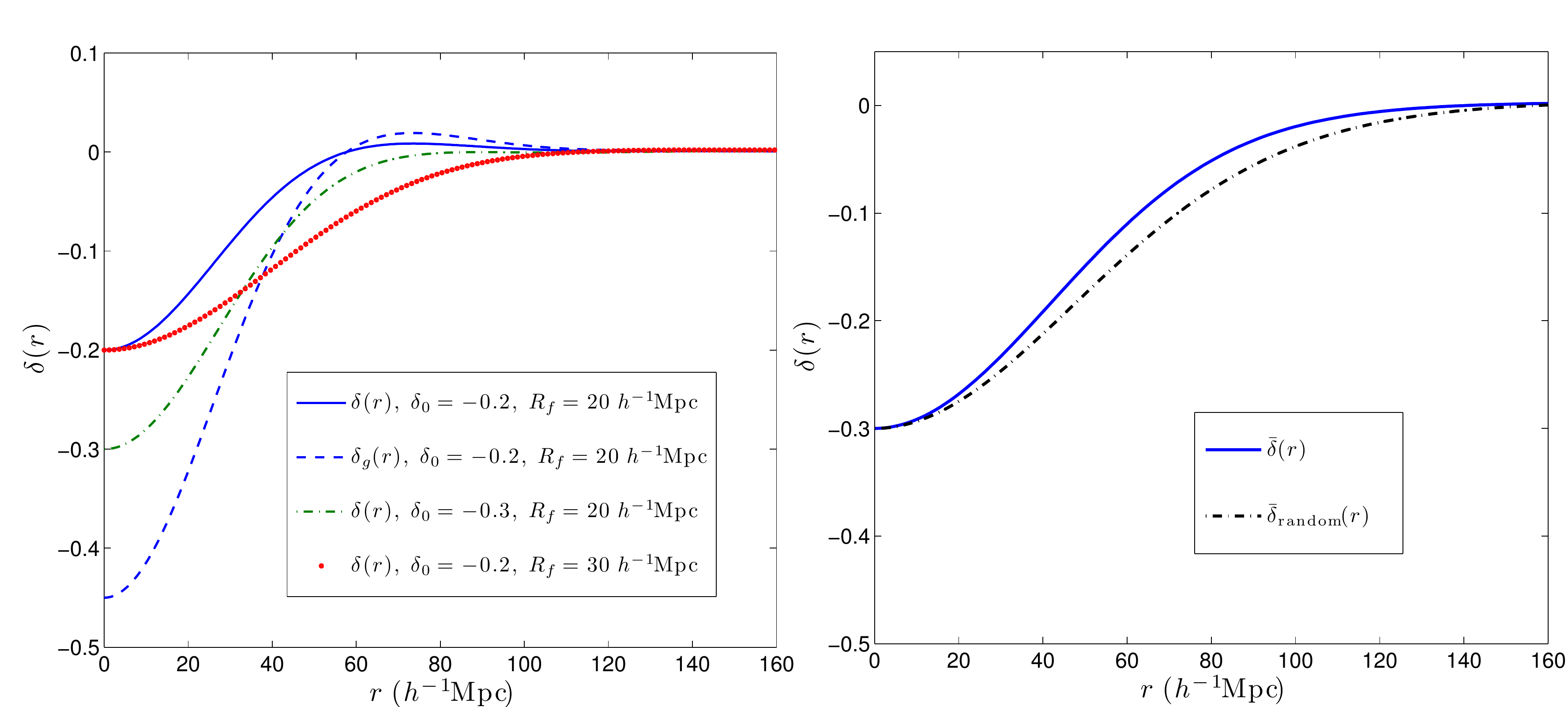} 
\caption{\label{fig:profiles} \emph{Left panel:} Mean radial profiles for voids obtained
  from Eq.~(\ref{eq:LiljeLahavform}) for different values of central
  underdensity $\delta_0$ and the smoothing scale $R_\mathrm{f}$. The
  blue (solid) line and the green (dash-dot) line are for the same
  $R_\mathrm{f}~(=20\;h^{-1}$Mpc) but different values of $\delta_0$,
  whereas the red (dotted) profile has the same $\delta_0~(=-0.2)$
  as the blue (solid) profiles, but a larger smoothing scale. The blue
  (dashed) curve is the biased galaxy density contrast corresponding
  to matter density contrast given by the blue (solid) line, with bias
  factor $b=2.25$ as is appropriate for LRGs.
  \emph{Right panel:} Comparison between the two methods of calculating mean 
  radial profiles. The black (dash-dot) line represents $\bar\delta_\mathrm{random}(r)$
  from Eq.~\eqref{eq:Dekel} for $\delta_0=-0.3$ and $R_\mathrm{f}=30\;h^{-1}$Mpc.
  The blue (solid) line is the corrected profile $\bar\delta(r)$ from 
  Eq.~\eqref{eq:LiljeLahavform} for the same parameter values.}
\end{figure}

We make the simplifying approximation that the average ISW signal for
a large number of superstructures is the same as the signal due to a
superstructure with the mean profile (\ref{eq:LiljeLahavform}) ---
this is justified because of the linear relationship between $\Delta
T$, $\Phi$ and $\delta$. The actual distribution of the size of
structures depends on the distribution of $x$ values, which can be
obtained from Eq. (7.5) of Ref.~\cite{Bardeen:1985tr}. In the left 
panel of Fig.~\ref{fig:profiles} we plot some underdense profiles for 
selected values of $\delta_0$ and $R_\mathrm{f}$ (these examples are 
chosen for clarity and are not representative of the most likely 
actual underdensities). Note that the size of the structures is much 
larger than the smoothing scale $R_\mathrm{f}$.

To identify superstructures in galaxy surveys (the methodology of 
Ref.~\cite{Granett:2008ju} is discussed in more detail in 
Section~\ref{section:comparison}) the galaxy density contrast 
$\delta_g$ is assumed to be linearly biased with
respect to the matter density: $\delta_g = b\delta$. Denoting by
$\rho_\mathrm{sl}$ the value of the density field at turnover, and by
$\rho_0$ the minimum density at the centre, a selection cut is made on
on $w \equiv \rho_\mathrm{sl}/\rho_0$, amounting to a lower bound on
the absolute value of $\delta_0$. This avoids false detections of
over- and under-dense regions that are just Poisson fluctuations in
galaxy number counts and ensures that only extreme superstructures are
included in the ensemble.

The definition of the radius $R_\mathrm{v}$ of any of the voids shown
in Fig.~\ref{fig:profiles} is slightly ambiguous. We choose it to be
the radius of turnover in the density profile less $R_\mathrm{f}$,
since smoothing necessarily increases the radius somewhat. This is a
small correction since in general $R_\mathrm{v}\gg R_\mathrm{f}$.

The mean profile about a \emph{random} point with a density contrast $\delta_0$
(i.e., a point which is not necessarily an extremum of the density field) is simply 
\cite{Dekel:1981}
\begin{equation}
\label{eq:Dekel}
\bar\delta_\mathrm{random}(r)=\delta_0\psi(r)\;.
\end{equation}
This should be compared with Eq.~\eqref{eq:BBKS7.10}, which is the appropriate
mean profile about points that are also known to be extrema. Eq.~\eqref{eq:Dekel}
is the form of the density profile used in Refs.~\cite{Papai:2010eu,Papai:2010gd}. 
In Ref.~\cite{Papai:2010gd} this is matched to the mean density profile about randomly 
chosen points in $N$-body simulations and is found to be a good approximation, as 
expected. It must be emphasised that as we identify superstructures with points 
of extrema in the density field, consistency demands that we use the corrected 
form of Eqs.~\eqref{eq:BBKS7.10} and~\eqref{eq:LiljeLahavform}
rather than Eq.~\eqref{eq:Dekel}. This gives a profile that is slightly steeper 
than that used in~\cite{Papai:2010eu,Papai:2010gd}.

In practical terms, however, the choice of profile between Eq.~\eqref{eq:BBKS7.10} 
and Eq.~\eqref{eq:Dekel} will have little effect on our conclusions as the difference 
between these profiles is small. In the right panel of Fig.~\ref{fig:profiles} we show as
an example the mean profiles $\bar\delta(r)$ and $\bar\delta_\mathrm{random}$ 
for $\delta_0=-0.3$ and $R_\mathrm{f}=30\,h^{-1}$Mpc. It can be seen that the two 
are similar: in fact, for most values of $\delta_0$ and $R_\mathrm{f}$ and over
almost the entire range of $r$, $\bar\delta_\mathrm{random}(r)$
lies within the $\pm1\sigma$ dispersion around $\bar\delta(r)$ of the profiles of 
extrema (this is not shown here but some examples can be seen in Fig. 8 of 
Ref.~\cite{Bardeen:1985tr}).

Of course the quantity determining $\Delta T$ in Eq.~\eqref{eq:SWintegral} is not 
$\delta(r)$ but an integral over $F(r)$, which is itself an integral over $\delta(r)$, and 
hence small differences may be magnified. Therefore, to explicitly check that the 
choice of profile is not responsible for the difference between our results 
and those of \cite{Papai:2010gd}, we also repeat the calculation of the expected 
temperature signal described below using $\bar\delta_\mathrm{random}(r)$
to model structures.

\subsection{Temperature signal}
\label{subsection:expectedtemp}

The ISW signal of any individual superstructure will be too small
compared to the primordial CMB anisotropies to be
observable. Therefore, what is measured is the average temperature
fluctuation along the lines of sight of a selected sample of either
over- or under-densities. The primordial anisotropies are uncorrelated
with the large scale structure and average out so given a large
enough sample the correlated ISW signal eventually dominates. Our
calculation of this averaged signal is done as follows.

At a given $z$, we use Eq.~(\ref{eq:LiljeLahavform}) to calculate the
matter and galaxy density profiles about extrema of the density field
as functions of $\delta_0$ and $R_\mathrm{f}$ and obtain $\dot\Phi$
along the line of sight as discussed in
Section~\ref{section:ISW}. This enables us to calculate $\Delta T
(\theta;\delta_0,R_\mathrm{f})$ where $\theta=0^\circ$ is the line of
sight passing through the centre of the superstructure. To compare
with the observations we first apply the selection criterion on
$\delta_0$ through the limit on $w$. Then, to calculate the
expectation value $\langle\Delta T\rangle$ for the resulting ensemble,
we weight the results appropriately with the number density 
(\ref{eq:diffnumberdensity}) of extrema. Hence for an ensemble of voids:
\begin{equation}
\label{eq:expectedtemp}
\langle\Delta T\rangle = \frac{\iint W(\theta)\Delta T (\theta;\delta_0) 
 \mathcal{N}_\mathrm{min}\sigma_0^{-1}\;\mathrm{d}^2{\bf \theta}d\delta_0}
 {\pi\theta_\mathrm{c}^2\int \mathcal{N}_\mathrm{min}\sigma_0^{-1}\; 
 \mathrm{d}\delta_0},
\end{equation}
where $0\leq\theta\leq\theta_\mathrm{out}$; $W(\theta)$ is a filter of 
angular width $\theta_\mathrm{out}$ chosen in order to match that used 
in the actual observation and $-1\leq\delta_0\leq\delta_0^\mathrm{c}$, 
where $\delta_0^\mathrm{c}$ is the (mildly $R_\mathrm{f}$-dependent) 
cutoff on the minimum underdensity required to pass the selection 
criterion. The choice of $R_\mathrm{f}$ determines the mean radial 
size of the voids included in the ensemble; although structure-finding 
algorithms may not have an explicit size dependence, in practice there 
is obviously a lower limit on the size of the over- or under-density 
that can be reliably found. As smaller structures are overwhelmingly 
more probable than larger ones, it is important to capture this effect 
and we discuss this in more detail in the next Section. For a given 
ensemble of voids specified by $R_\mathrm{f}$ and 
$\delta_0^\mathrm{c}$, the observed temperature signals $\Delta T$ 
will be distributed about this mean value. By calculating 
$\langle\Delta T^2\rangle$ in a similar fashion to Eq.~(\ref{eq:expectedtemp}) 
we can estimate the standard deviation of this distribution simply 
as $\left(\langle\Delta T^2\rangle-\langle\Delta T\rangle^2\right)^{1/2}$.

The expected signal from an ensemble of clusters follows in an
exactly analogous manner to Eq.~(\ref{eq:expectedtemp}).

\section{Comparing theory to observation}

\subsection{The measured ISW signal of superstructures}
\label{section:observation}

To compare our expectation for the ISW signal to the measurement made
by Ref.~\cite{Granett:2008ju}, it is necessary to follow the same selection procedure. They
use a sample of 1.1 million LRGs in the range $0.4<z<0.75$ (median
$z=0.52$) from the SDSS DR6 \cite{Adelman-McCarthy:2007wu}, which
covers 7500 degree$^2$ on the sky and occupies a volume of
$5\;h^{-3}\mathrm{Gpc}^3$. They search for `supervoids' and
`superclusters' using two publicly-available structure-finding
algorithms: {\small ZOBOV} (ZOnes Bordering On Voidness;~\cite{Neyrinck:2007gy}) for supervoids, and {\small VOBOZ} (VOronoi
BOund Zones;~\cite{Neyrinck:2004gj}) for superclusters.

It is necessary to mimic the way in which these algorithms
select structures in choosing the ensemble for which to calculate
$\langle\Delta T\rangle$ from Eq.~(\ref{eq:expectedtemp}).  {\small
ZOBOV} uses a parameter-free Voronoi tessellation to estimate the
density at each galaxy in the sample, based on the distance to its
nearest neighbours. Around each density minimum it then finds the
region of the density depression or supervoid. (Of course large voids
can contain multiple smaller voids, or even isolated high-density
regions.) The `significance' of the depression is estimated by
comparing the density contrast, $w$ (defined as the ratio of the
density at the lip of the void to the density at its minimum) to an
uniform Poisson point sample. This yields the likelihood that a void
of density contrast $w$ could arise from Poisson noise, i.e. that it
is a false positive detection; a $3\sigma$ cut is then applied on the
likelihood which translates to requiring $w>w_\mathrm{c}=2.0$ on the
density contrast \cite{Neyrinck:2007gy}. This procedure yields 50
supervoids, the properties of which are tabulated in
Ref.~\cite{Granett:2008xb}. The {\small VOBOZ}
supercluster finder uses the same algorithm but applied to the inverse
of the density field, with density contrast defined as the ratio of
the peak density to the density at the edge of the structure. However,
now the $3\sigma$ cut on the likelihood that an overdensity of given
$w$ could have arisen due to Poisson noise corresponds to
$w>6.8$~\cite{Neyrinck:2004gj}. In fact the authors impose a tighter cut:
$w>8.35$ in order to obtain exactly 50 such superclusters; their
properties are also tabulated in Ref. \cite{Granett:2008xb}.

Ref. \cite{Granett:2008ju} report a search for the ISW signals of these 
superstructures using an inverse-variance weighted combination 
of the WMAP 5-year Q, V and W maps~\cite{Hinshaw:2008kr} with 
foreground subtracted and the KQ75 mask applied. They build stacked 
images by averaging the CMB temperature in the regions around the 
lines of sight passing through the centres of the identified superstructures 
and use a compensated top-hat filter of width $\theta_\mathrm{c}$ 
in order to perform the averaging. This corresponds to making the choice in
Eq.~(\ref{eq:expectedtemp}):
\begin{eqnarray}
\label{eq:tophatfilter}
W(\theta) = \left\{
\begin{array}{cr}
  \displaystyle 1\,,   &  0\leq\theta\leq\theta_\mathrm{c}; \\
  \displaystyle -1\,,  & \theta_\mathrm{c}<\theta\leq\theta_\mathrm{out},
\end{array}
\right.
\end{eqnarray}
with $\theta_\mathrm{out}=\sqrt{2}\theta_\mathrm{c}$. For a filter
radius $\theta_\mathrm{c}=4^\circ$,\footnote{This choice is 
apparently motivated by the expectation that the CMB-galaxy 
cross-correlation should peak at about $4^\circ$~\cite{Padmanabhan:2004fy}. 
Ref. \cite{Granett:2008ju} reports repeating the observation with a 
few other widths, $3^\circ\leq\theta_\mathrm{c}\leq5^\circ$ and 
obtaining a maximum detection significance for $\theta_\mathrm{c}=4^\circ$.} 
the sample of supervoids gives $\langle\Delta T\rangle=-11.3\pm3.1\;\mu$K, 
and the sample of superclusters $\langle\Delta T\rangle=7.9\pm3.1\;\mu$K. 
When averaged together (with the negative of the supervoid image added 
to the superclusters) this gives $\langle\Delta T\rangle=-9.6\pm2.2\;\mu$K, 
i.e. a $4.4\sigma$ detection of a non-zero correlation. 

The quoted error bars are determined by Monte Carlo simulations in 
two different ways: by stacking images of the actual CMB map in random 
directions, and by keeping the directions fixed but using model CMB maps. 
Note that this explicitly accounts for cosmic variance in the CMB fluctuations. 
The theoretical error bars described in Section~\ref{subsection:expectedtemp} 
account for the distribution of voids of different depths and sizes. Misidentification 
of structures in the dark matter density field can also occur due to shot 
noise in the galaxy distribution. This means that the structures identified as 
the $N$ most extreme fluctuations will in fact be a random sample of $N$ 
structures drawn from among the $M$ most extreme, for some $M\geq N$. In 
Section~\ref{subsection:selection}, we make the optimistic assumption 
that $N=M$ such that the expected signal is maximised.

A caveat that should be noted, however, is that the choice in Ref.~
\cite{Granett:2008ju} of using exactly $50$ supervoids and superclusters 
each is made to obtain the largest signal-to-noise ratio. For both $N=30$ 
and $N=70$ a somewhat lower significance detection is reported compared 
to $N=50$. As the distribution of $\Delta T$ values for the voids is very 
non-Gaussian, this choice of $N$ could be important. Fluctuations due to 
the underlying CMB anisotropy dominate at small $N$ and false positive 
identifications of structures increase at large $N$ so the signal-to-noise 
ratio is indeed expected to have a maximum at some intermediate $N$. 
However more detailed study is needed to quantify whether this is expected 
to be at $N\simeq50$. In this paper we take the reported observations at face 
value and our interpretation is based on the data for $N=50$. 

\subsection{Comparison to the theoretical expectation}
\label{section:comparison}

We assume for simplicity that all the superstructures are located at
the mean redshift $z=0.52$ and adopt the standard $\Lambda$CDM
cosmological model with $\Omega_\mathrm{m}=0.29$,
$\Omega_\Lambda=0.71$, $n_\mathrm{s}=0.96$ and $\sigma_8=0.83$ (with
$h=0.69$ where required) to obtain the matter power spectrum at
$z=0.52$ using CAMB \cite{Lewis:1999bs}.\footnote{These are the mean
parameter values obtained from a fit to WMAP
7-year~\cite{Komatsu:2010fb} and SDSS DR7~\cite{Abazajian:2008wr}
data using COSMOMC~\cite{Lewis:2002ah}.}  The bias factor for LRGs
is taken to be $b=2.25$. These parameters are kept fixed for the purposes
of calculation, but as we discuss later, varying them within the bounds set by
other observational data has no effect on the conclusions.

Our first finding is that there are no \emph{overdense}
superstructures within the linear regime (i.e. with $\delta_0 <1$)
which meet the {\small VOBOZ} $3\sigma$-significance
selection criterion that $w>6.82$. Such a ratio of densities (between
the lip of an overdensity and its centre) can be achieved, but only for
\emph{non-linear} collapsed structures. We conclude that Table 5 of
Ref. \cite{Granett:2008xb} does not list the most overdense large-scale linear
perturbations, but the (mild) large-scale linear perturbations that
happen to contain the most overdense small-scale collapsed
structures. This means that the relative ranking and selection of 
large-scale linear overdensities made in Ref.~\cite{Granett:2008xb} 
is done on the basis of the small-scale collapsed structures they 
contain. The statistical effect on the CMB temperature of small 
collapsed structures could be calculated, but we cannot model the criteria by 
which these overdense regions were selected for study. Note however that
our calculations below for the maximum possible amplitude of ISW
signal from superstructures holds equally well for over- and
under-densities in the linear regime. Any contamination of the regions
selected by {\small VOBOZ} will only reduce the expected signal,
although we are unable to estimate by how much.

On the other hand, this problem of modelling the selection of 
structures for study does not arise 
for voids. Hence we concentrate only on the sample of \emph{underdense} 
regions (i.e. supervoids), for which $\langle\Delta T\rangle_\mathrm{obs} =
-11.3\pm3.1\;\mu\mathrm{K}$. To calculate the expected $\langle\Delta
T\rangle$ using Eq.~(\ref{eq:expectedtemp}), we must first choose the
smoothing scale $R_\mathrm{f}$ and determine the distribution of void
radii $R_\mathrm{v}$ in the ensemble using Eq.~(7.5) of 
Ref.~\cite{Bardeen:1985tr}. For each
$R_\mathrm{f}$ we calculate $R_\mathrm{v}^\mathrm{min}$ such that 95\%
of all voids in the ensemble have radius
$R_\mathrm{v}>R_\mathrm{v}^\mathrm{min}$, this being a convenient way
to characterise the ensemble. For $R_\mathrm{f}=20\;h^{-1}$Mpc, we
find $R_\mathrm{v}^\mathrm{min}\sim70\;h^{-1}$Mpc, which is similar to
the mean radius of the voids in Table 4 of Ref. \cite{Granett:2008xb};\footnote{At 
the radii reported here, the density has not yet reached the background
level so the quoted values must be underestimates of the void size
relative to our criterion.} at this scale, the number of such voids
that should satisfy the selection criterion is
$N_\mathrm{v}\sim10^4$. This is to be compared with the
$N_\mathrm{v}=50$ voids that are actually tabulated in Ref. \cite{Granett:2008xb}. 
For the larger ensemble, we find an expectation value $\langle\Delta T\rangle
=-0.3\pm0.2\;\mu\mathrm{K}$, i.e. consistent with zero.

However, since Granett \emph{et al.} see only a tiny fraction of the 
total number of supervoids expected in the SDSS DR6 volume, it is 
unreasonable to assume that this represents a fair sample. Some strong 
selection effects \emph{must} be in operation; while these have no effect on 
our ISW prediction for individual voids, they will affect the selection of 
structures in the LRG distribution and thus the ensemble expectation value. 
Selection effects that can enhance the expected signal are a 
bias towards larger and deeper regions, implying that Granett \emph{et al.} 
did not randomly select 50 of the $\sim10^4$ expected supervoids but chose
some sample that is skewed towards regions with larger $\Delta T$
values. We show below that the expected signal from the 50 most
extreme regions is indeed $\sim5$ times larger than $-0.3\;\mu
\rm{K}$; nevertheless the discrepancy with the observation in 
Ref.~\cite{Granett:2008ju} is still $>3\sigma$. Therefore, irrespective of how 
the regions were selected, tension remains with the expectation in 
the standard $\Lambda$CDM model.

\begin{figure*}
\center
\includegraphics[width=0.8\columnwidth]{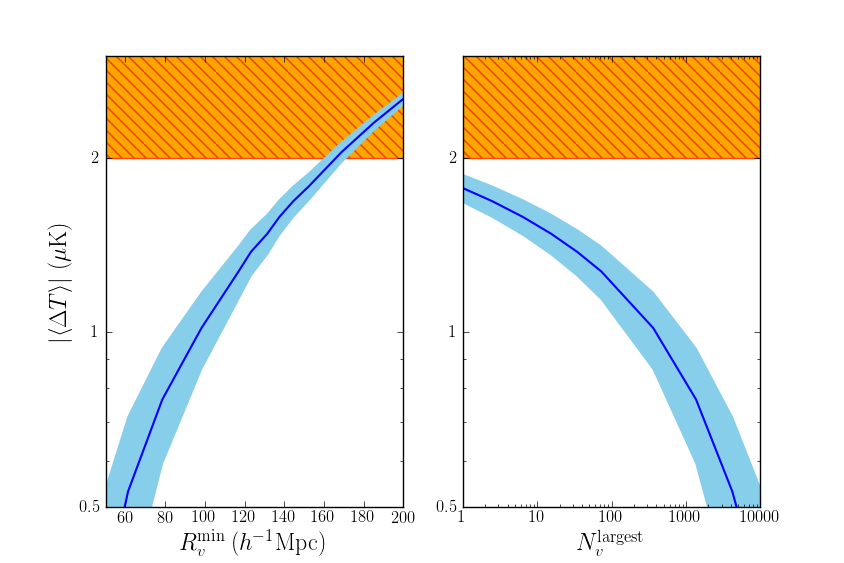}
\caption{\label{fig:Tlargest} \emph{Left panel}: The absolute value of
  $\langle\Delta T\rangle$ for an ensemble of voids which satisfy the
  {\small ZOBOV} selection condition on density (see text), as a
  function of the minimum radius of voids in the ensemble. The solid
  (blue) curve shows the mean value and the shaded (lighter blue)
  contours the $1\sigma$ region. The (orange) cross-hatched area is
  the lower end of the $3\sigma$ range of the observed value
  $\langle\Delta T\rangle_\mathrm{obs} = -11.3\pm3.1\;\mu\mathrm{K}$.
  \emph{Right panel}: As above, but showing $\langle\Delta T\rangle$
  as a function of the number of voids in the ensemble from which the
  observed sample of 50 voids is to be drawn, when only the
  $N_\mathrm{v}$ \emph{largest} voids also meeting the {\small ZOBOV}
  selection condition on density are included.}
\end{figure*}

\subsubsection{Accounting for selection effects}
\label{subsection:selection}

We first consider the possibility that the void-finding algorithm
{\small ZOBOV} is sensitive only to the largest (and least common)
voids in the matter distribution which produce the biggest ISW
temperature signals. The average density of LRGs in the SDSS DR6 
sample is roughly 1 galaxy per $(15\;h^{-1}\mathrm{Mpc})^3$. In 
underdense regions LRGs will be even more sparsely distributed 
so {\small ZOBOV} will certainly be less able to identify smaller 
underdense structures, thus biasing the sample towards larger voids.

In order to model the effect of such a selection bias, we increase the
value of $R_\mathrm{f}$ in Eq.~(\ref{eq:expectedtemp}); this is
equivalent to including only the $N_\mathrm{v}$ \emph{largest} voids
with radius $R_\mathrm{v}\geq R_\mathrm{v}^\mathrm{min}$ in the
ensemble from which $\langle\Delta T\rangle$ is calculated. In the
left panel of Fig.~\ref{fig:Tlargest} we plot the expectation
$\langle\Delta T\rangle$ as a function of
$R_\mathrm{v}^\mathrm{min}$. The orange cross-hatched area shows the
region that is within $3\sigma$ of the observed value $\langle\Delta
T\rangle_\mathrm{obs} = -11.3\pm3.1\;\mu\mathrm{K}$. The theoretical
value of $\langle\Delta T\rangle$ becomes marginally consistent with
the observed value when
$R_\mathrm{v}^\mathrm{min}\sim170\;h^{-1}$Mpc. However the probability
that within the SDSS survey volume there are 50 supervoids of radius
$R_\mathrm{v}\geq170\;h^{-1}$Mpc which also meet the {\small ZOBOV}
selection criterion is vanishingly small.

It is interesting to note that the 50 largest supervoids expected
within the SDSS survey volume have
$R_\mathrm{v}^\mathrm{min}\gtrsim120\;h^{-1}$Mpc. The largest void
radius reported in Ref.~\cite{Granett:2008xb} is $125\;h^{-1}$Mpc and 
the mean is $70\;h^{-1}$Mpc. We have argued that these values somewhat
underestimate the size of the supervoids compared to our definition of
$R_\mathrm{v}$, yet it seems unlikely that the difference could be so
large that \emph{all} the tabulated voids should have
$R_\mathrm{v}\gtrsim120\;h^{-1}$Mpc.

In the right-hand panel of Fig.~\ref{fig:Tlargest} we plot
$\langle\Delta T\rangle$ as a function of the size $N_\mathrm{v}$ of
the ensemble of the largest supervoids that should exist within the
SDSS volume. It is from this ensemble that the 50 observed supervoids
should be regarded as having been drawn. It can be seen that even
under the assumption that the {\small VOBOZ} algorithm
selected exactly the 50 largest voids in the entire SDSS survey
volume, the expected signal is only $\langle\Delta
T\rangle=-1.33\pm0.13\;\mu$K which is still discrepant by $>3\sigma$
with the observed value. We conclude that the observed signal cannot
be explained due to a simple bias towards selecting only the largest
voids.

We repeated the calculation above using Eq.~\eqref{eq:Dekel} to model 
the density profiles of the voids, as in Ref.~\cite{Papai:2010gd}, and 
obtained $\langle\Delta T\rangle=-1.58\pm0.12\;\mu$K for the 50 largest voids. Despite
the small increase compared to the value quoted above, this is still discrepant 
with observation at the same $>3\sigma$ level. This conclusively demonstrates 
that, as anticipated, the difference between our conclusion and that of 
Ref.~\cite{Papai:2010gd} is not due to the small correction included in $\bar\delta(r)$. 
Henceforth we use the corrected profile only.

\begin{figure*}
\center
\includegraphics[width=\columnwidth]{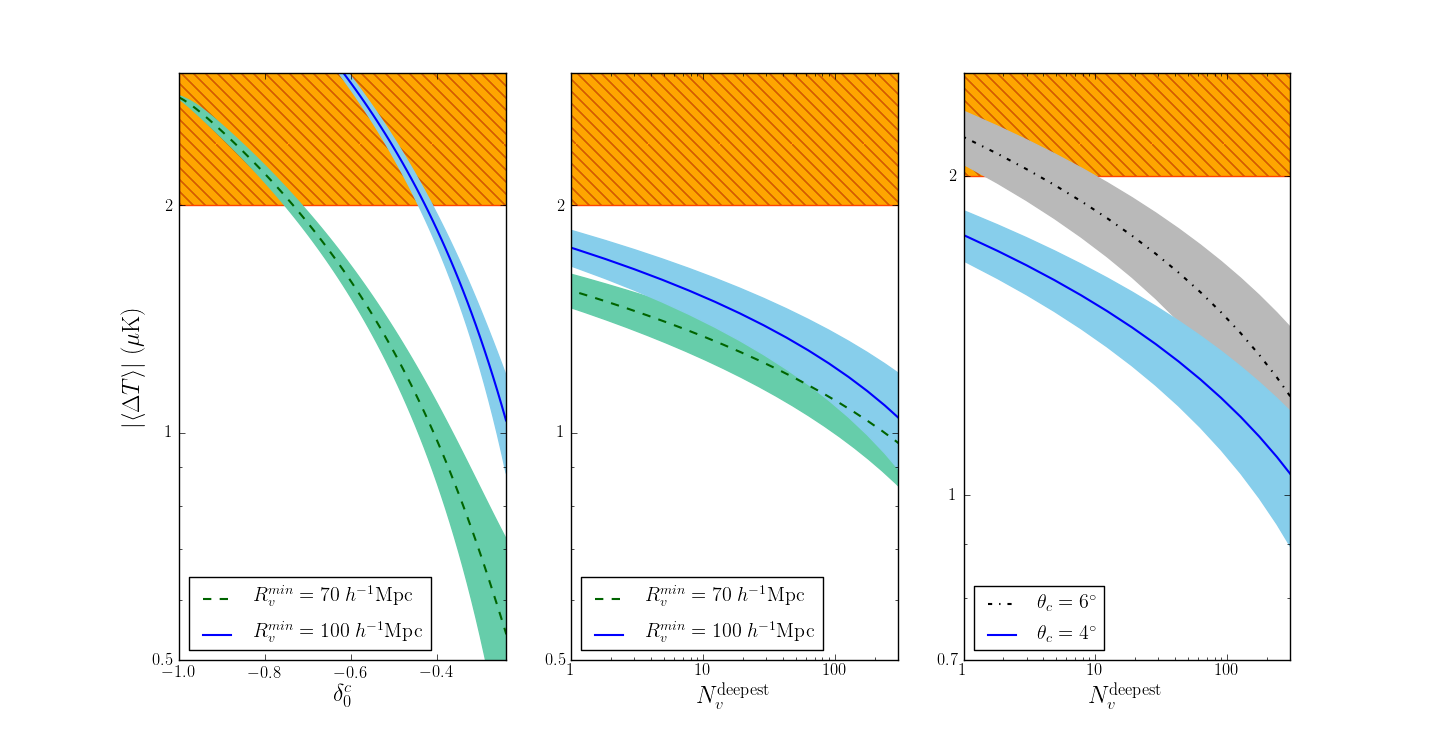}
\caption{\label{fig:Tdeepest} \emph{Left panel}: The absolute value of
  $\langle\Delta T\rangle$ for an ensemble of supervoids which satisfy
  $\delta_0<\delta_0^\mathrm{c}$, as a function of $\delta_0^\mathrm{c}$. The dashed
  (green) curve shows the case when $R_\mathrm{f}$ is chosen such that
  voids with $R_\mathrm{v}\gtrsim70\;h^{-1}$Mpc are included in the
  ensemble; the solid (blue) curve is for
  $R_\mathrm{v}\gtrsim100\;h^{-1}$Mpc. Shaded contours show the
  $1\sigma$ region about the mean and the orange cross-hatched area is
  as in Fig.~\ref{fig:Tlargest}. \emph{Middle panel}: As before, but
  showing $\langle\Delta T\rangle$ as a function of the number of
  voids in the ensemble from which the observed sample of 50
  supervoids is to be drawn, when only the $N_\mathrm{v}$
  \emph{deepest} supervoids are included. \emph{Right panel}: As
  before, but for two different choices of the radius
  $\theta_\mathrm{c}$ of the compensating top-hat filter used in
  Eq.~(\ref{eq:expectedtemp}). The solid (blue) curve is the mean
  value for $\theta_\mathrm{c}=4^\circ$, and the dot-dashed (black) curve
  for $\theta_\mathrm{c}=6^\circ$. Shaded regions show the $1\sigma$
  deviations from the mean.}
\end{figure*}

We consider next whether the {\small ZOBOV} algorithm is more
sensitive to deeper voids. In Table 4 of Ref.~\cite{Granett:2008xb}, the edge 
of most of the supervoids is defined at a radius where the density contrast is still
negative. This means {\small ZOBOV} systematically underestimates the
value of $w$ relative to our definition (where
$\delta_{\rm{edge}}\simeq0$), so Granett \emph{et al.} effectively
used a more stringent cut on $w$ than we have done. We can
model this effect by varying $\delta_0^\mathrm{c}$ from the value determined by
the stated algorithm. In the left-hand panel of
Fig.~\ref{fig:Tdeepest} we plot as examples $\langle\Delta T\rangle$
as a function of $\delta_0^\mathrm{c}$ for
$R_\mathrm{v}^\mathrm{min}\simeq70\;h^{-1}$Mpc (the mean radius of the
supervoids in \cite{Granett:2008ju}) and
$R_\mathrm{v}^\mathrm{min}\simeq100\;h^{-1}$Mpc. The right-hand panel
shows $\langle\Delta T\rangle$ as a function of $N_\mathrm{v}$, the
number of voids included in the ensemble when $\delta_0^\mathrm{c}$ is varied
in each case. For the smaller radius, the ensemble is dominated by
small voids with a small ISW effect, so increasing
$R_\mathrm{v}^\mathrm{min}$ increases $\langle\Delta T\rangle$ at any
$\delta_0^\mathrm{c}$. With $R_\mathrm{v}^\mathrm{min}\sim100\;h^{-1}$Mpc,
$\langle\Delta T\rangle$ becomes marginally consistent with the
observation for $\delta_0^\mathrm{c}\lesssim-0.5$. However, the probability of
obtaining 50 supervoids with $R_\mathrm{v}\geq100\;h^{-1}$Mpc and
$\delta_0\leq-0.5$ is negligibly small, as the right-hand panel
clearly demonstrates. Therefore we conclude that in the standard 
cosmological picture there do not exist 50 isolated matter density voids 
in the SDSS volume that, irrespective of the method of selecting them, 
produce an average $\langle\Delta T\rangle$ compatible with observation. 
Our analysis, based on very conservative assumptions, shows that the claim 
\cite{Papai:2010gd} that the discrepancy between observation and the 
$\Lambda$CDM prediction is only $\sim2\sigma$ is an underestimate. 

\subsubsection{Potential systematic errors}
\label{section:discussion}

We now discuss the expected corrections due to the simplifying
assumptions we made in calculating the signal. We first consider the 
effect of varying the cosmological parameters from the 
mean values adopted in \ref{section:comparison}. This can affect the calculation
of $\langle\Delta T\rangle$ in three ways:\footnote{Varying the bias factor only 
affects the detection of matter density voids via the observation of galaxies. By varying
$\delta_0^\mathrm{c}$ as above we have already made the conservative assumption that 
only the most extreme matter voids that exist are detected. Therefore the 
bias factor can have no further effect on our conclusions.} by altering the mean 
radial profile of the voids; by altering the expected number density of voids through 
Eq.~\eqref{eq:diffnumberdensity}; and by altering the ISW growth factor $G(z)$ 
and the integral Eq.~\eqref{eq:SWintegral}. The first two of these are dependent, 
through the moments $\sigma_i$, on the matter power spectrum $P(k)$. An increase 
in power increases the number density of points of extrema, and steepens the mean 
profile about these points. However, it also increases $\sigma_8$, which is already 
reasonably well constrained by WMAP \cite{Komatsu:2010fb} and SDSS DR7~
\cite{Abazajian:2008wr} data. Within any reasonable range of $\sigma_8$, the change 
to the mean profile $\bar\delta(r)$ is completely negligible. Allowing $\sigma_8=0.87$, 
which is the $95\%$ C.L. bound from WMAP + SDSS, does boost the relative
number density of the largest voids by a factor of $\sim1.5$. However, given the
almost logarithmic dependence of $\langle\Delta T\rangle$ on 
$N_\mathrm{v}^\mathrm{largest}$ evident from Fig.~\ref{fig:Tlargest}, the effect on 
$\langle\Delta T\rangle$ remains well within the statistical error bars already quoted.

The only remaining way for the cosmological parameters to affect $\langle\Delta T\rangle$
is through the value of $\Omega_\mathrm{m}$. Assuming a spatially flat model,
this determines the growth factor $G(z)$, the constant pre-factor in 
$\dot\Phi$ and the photon geodesics (i.e. the relationship between $r$ and $z$ in 
Eq.~\eqref{eq:SWintegral}). Note that a smaller $\Omega_\mathrm{m}$ boosts $G(z)$
but decreases the pre-factor. $\Omega_\mathrm{m}$ is of course also constrained
by other data, and we find that any value in the range $\Omega_\mathrm{m}\in(0.25,\,0.32)$ 
results in changes to $\langle\Delta T\rangle$ that are significantly smaller than 
the error bars already quoted. We conclude that the freedom to vary the $\Lambda$CDM 
model parameters is not enough to overcome the discrepancy with observation.

We have neglected non-linear effects and the time evolution of the void density 
profile \cite{Sheth:2003py,Colberg:2004nd} in the calculation of 
$\langle\Delta T\rangle$. On small scales voids evolve non-linearly towards a 
compensated top-hat profile. These non-linear effects increase 
the depth of the cold spot at its centre but also produce a hot filamentary 
shell~\cite{Cai:2010hx}. When averaged over angle $\theta$ it is not clear whether 
the net effect will lead to a net increase or decrease in $\langle\Delta T\rangle$. 
In fact, highly non-linear void profiles are known to produce a \emph{smaller} 
ISW signal due to the effect of the overdense ridge at the boundary
\cite{Inoue:2010rp,Papai:2010eu}, so by using the linear theory
profile we may be \emph{over}estimating the expected signal. It must be 
stressed though that at small redshifts $z\sim0.5$ and on the large scales 
of interest $\gtrsim100\;h^{-1}$Mpc, the effects of non-linear evolution will 
in any case represent only a subdominant correction \cite{Cai:2010hx} and can 
justifiably be neglected here.

A more subtle assumption is that the real supervoid profiles are adequately
captured by our smoothing prescription. The effect of the smoothing is to
slightly broaden the $\delta(r)$ profile which in turn leads to broadening of
the $\Delta T(\theta)$ profile. This may become a problem if the $\Delta
T(\theta)$ profile is significantly broadened so the compensated top-hat filter
of radius $4^\circ$ in Eq.~(\ref{eq:expectedtemp}) then underestimates the real
signal. If this were the case then it would be more appropriate to use a broader
top-hat. To check this, we repeated our analysis with a filter of radius
$6^\circ$, which is a generous overestimate of the degree of broadening, and in
the right panel of Fig. \ref{fig:Tdeepest} we show the effect for the largest
and deepest supervoids. As expected, increasing the filter radius does increase
$\langle\Delta T\rangle$, but the effect is small. Even with a $6^\circ$ filter
the expected signal remains $>3\sigma$ discrepant with observations for
$N_\mathrm{v}=50$; also as mentioned above, the actual effect on $\langle\Delta
T\rangle$ due to the smoothing will be less than this extreme model. This 
indicates that our conclusion of a discrepancy will be independent of the details
of the smoothing filter used.

We have made the simplifying assumption that the centres of all the 
superstructures lie at the same redshift $z=0.52$, which is the median 
redshift of the DR6 survey. This means we have not accounted for the 
possibility that two voids may lie at different redshift along
roughly the same line of sight, thus increasing the temperature shift. 
Equally we do not account for the possibility that a void may lie in front 
of or behind an overdensity, thus \emph{decreasing} the temperature shift. 
We expect that these two simplifications largely negate each other. 

As a test of the robustness of our calculation, we compare our
results with Fig.~1 of Ref.~\cite{Cai:2010hx} which shows the ISW map from
a cosmological $N$-body simulation for a volume comparable to the SDSS
DR6, but at $z=0$ rather than $z=0.52$. The \emph{very} largest
density perturbations in this map yields a maximum ISW signal of
$\vert\Delta T\vert\sim4\;\mu$K before applying a filter
analysis. Taking into account that $\Delta T$ is more pronounced at
smaller redshift, this tallies very well with our prediction from
Fig.~\ref{fig:Tdeepest} that the most extreme supervoid in the SDSS
volume should produce $\langle\Delta
T\rangle\sim-2\;\mu$K. 

As seen in the full-sky maps in Ref.~\cite{Cai:2010hx}, there are lines of
sight along which the chance alignment of several large structures
between us and the LSS several Gpc away can lead to isolated ISW 
`cold spots' with $\Delta T<-10\;\mu$K. However, this is a cumulative 
effect of several structures over Gpc scales and does not affect our 
conclusions about the correlation between these cold spots and isolated 
structures at $z\sim0.5$. (Note also that in the simulation the temperature profiles
along such lines of sight are much broader that $4^\circ$ and so when 
averaged using a $4^\circ$ compensated top-hat filter would still give only a 
small $\vert\langle\Delta T\rangle\vert$.) Similarly, there will be several $4^\circ$ 
circles on the sky for which the underlying CMB anisotropy alone can give an
average $\Delta T$ of similar magnitude --- it is just such
fluctuations that generate the observational uncertainty of
$\pm3.1\;\mu$K of Ref.~\cite{Granett:2008ju}. Such fluctuations are not due to any ISW 
effect and there is no reason why they should be correlated with large 
structures in the SDSS catalogue. Therefore, assuming only that the lines of 
sight were not chosen \emph{a posteriori}, the observed correlation between 
galaxy structures and the CMB can only result from rare ($>3\sigma$) fluctuations 
or anomalously large density perturbations at $z\sim0.5$.

Ref.~\cite{Granett:2008ju} also reports the variation of the signal-to-noise 
ratio with the width of the compensating top-hat filter and finds a maximum at
$\theta_\mathrm{c}=4^\circ$ for the combined sample of over-and
under-dense structures. By contrast, for supervoids we find that
$\vert\langle\Delta T\rangle\vert$ increases as $\theta_\mathrm{c}$ is
raised from $4^\circ$ to $6^\circ$ (Fig.~\ref{fig:Tdeepest}) and to larger angles. 
This is incompatible with a peak in the signal-to-noise ratio at $4^\circ$ and is
another respect in which the observation of Ref.~\cite{Granett:2008ju} 
cannot be modelled by linear structures in $\Lambda$CDM.

\subsection{Comparison with earlier results}
\label{subsection:differences}

We have provided a theoretical calculation of the expected average temperature 
signal in the CMB from superstructures observed in the SDSS LRG catalogue 
assuming a $\Lambda$CDM cosmology with gaussian primordial density 
perturbations. On comparing this expectation to the observation made in 
Ref.~\cite{Granett:2008ju}, we find a discrepancy of $>3\sigma$, even if 
possible unknown selection effects are accounted for. In fact we find that 
the \emph{single most extreme} void expected within the SDSS survey 
volume does not produce an angle-averaged $\Delta T$ value within $3\sigma$ of the 
observed average signal for 50 voids in G08a. On the other hand, in Ref.
\cite{Papai:2010gd} Papai \emph{et al.} perform a similar comparison using a 
different method, and report a discrepancy with $\Lambda$CDM of only 
$\sim2\sigma$. As demonstrated in Section~\ref{subsection:selection}, 
this is \emph{not} due to the slightly different form of the radial profile for 
superstructures used. Instead the source of the difference arises from the 
interpretation of the matched filter technique employed in Ref. \cite{Papai:2010gd}. 

Papai \emph{et al.} calibrate their radial profiles against simulation to obtain a template 
for the ISW effect from structures. As they note, reconstructing accurate estimates 
of the ISW profiles from comparison with simulations of finite size is complicated 
by cosmic variance in the simulation, necessitating the use of a high-pass filter in 
the reconstruction. Once this filter is applied, their numerical template profiles agree 
well with our analytic ones.

They then assume exactly the same template profile for all structures, scaled  
by choosing $\delta_0=1$ for superclusters and $\delta_0=-1$ for supervoids, 
and place these structures at redshift $z=0.52$ at the locations  of the structures 
observed in G08a. However, this procedure does not account for the 
actual density contrast of structures that can be expected
in $\Lambda$CDM with gaussian primordial perturbations, nor does it 
account for the fact that not all structures will be identical. 
By contrast our analysis models the actual expected distribution of void 
depths and sizes. Although, as in our analysis, all structures are assumed to 
be centred at the same redshift for the purposes of calculating the template ISW 
shift for individual structures, when lines of sight overlap their template contributions 
are simply added. 

In this way Papai \emph{et al.} obtain the template ISW temperature map, scale 
this template map by a factor $\lambda$, and maximise the likelihood function
\begin{equation}
L(\lambda)=-\frac{1}{2}\left(T_\mathrm{CMB}-\lambda 
T_\mathrm{ISW}\right)C^{-1}\left(T_\mathrm{CMB}-
\lambda T_\mathrm{ISW}\right)\;,
\end{equation}
where $T_\mathrm{CMB}$ is the WMAP ILC map, $T_\mathrm{ISW}$ is the 
constructed template ISW map and $C$ is the pixel covariance matrix 
obtained from best-fit $\Lambda$CDM. 

The value of $\lambda$ that maximises the likelihood 
$L(\lambda)$ can be interpreted as the average central density contrast 
in the superstructures \cite{Papai:2010gd}
\begin{equation}
\lambda\approx\frac{1}{2}\left(\delta_0^{50c}-\delta_0^{50v}\right)\;,
\end{equation}
where superscripts $50c$ and $50v$ refer to the average of $50$ 
superclusters and $50$ supervoids respectively (note that in the linear 
approximation voids and clusters are entirely equivalent except for a minus sign). 
This is merely the density contrast that the superstructures must have in order for 
the template to fit the data. 

It is clear from the figures in Ref.~\cite{Papai:2010gd} that the best-fit values 
are $\lambda>1$ for a wide range of radial scales of the structures. Thus, although
the initial scenario, where all structures are postulated to have exactly 
$|\delta_0|=1$, is already extremely unlikely, in order to fit the observed 
CMB the template ISW map must be further scaled up by a factor $\lambda>1$. 
In this case the linear approximation used to calculate the template ISW map 
must break down (indeed it is not clear that the approximation can be valid even 
when $|\delta_0|=1$). Further note that physical voids are restricted to 
$\delta_0>-1$ making the interpretation of $\lambda>1$ even more 
problematic. Therefore although Ref.~\cite{Papai:2010gd} find a reasonable 
fit for their template when $\lambda>1$, this value is indicative of a problem 
with modelling the observations. This qualitative conclusion is entirely 
compatible with the more careful quantitative calculation presented here. 

Although this conclusion is not highlighted in Ref.~\cite{Papai:2010gd} it is clear 
that correctly interpreted, both results in fact indicate a significant tension 
between the $\Lambda$CDM model of the ISW effect due to 
superstructures and the observation first reported in Ref.~\cite{Granett:2008ju}.

\section{Summary and prospectives}
\label{section:conclusions}

We have calculated the integrated Sachs-Wolfe effect expected in
$\Lambda$CDM from $\sim 100\;h^{-1}$Mpc size structures, using the
density profiles predicted by the linear theory of gaussian
perturbations~\cite{Bardeen:1985tr}. We find that the most extreme
superstructures in the SDSS volume will produce an ISW signal of $\lesssim
2\;\mu$K. This matches well with ISW maps generated from $N$-body
simulations of the $\Lambda$CDM cosmology \cite{Cai:2010hx}.

Our result is about 3 times larger than earlier calculations 
\cite{Hunt:2008wp,Inoue:2010rp} which assumed compensated 
top-hat density profiles. Such an assumption, while well-motivated 
for non-linear structures that have formed at small scales 
\cite{Sheth:2003py}, should not apply on the large scales of the 
superstructures considered here. Ref.~\cite{Papai:2010eu} noted that 
the ISW signal from such structures should therefore be larger and 
our results confirm this.

Nevertheless we have demonstrated that the ISW signal claimed to have
been detected in Ref.~\cite{Granett:2008ju} is still $>3\sigma$ larger
than the signal expected in $\Lambda$CDM. This tension persists even
after allowing for likely selection effects. In fact, even the most
extreme underdensities in the SDSS volume would still produce a signal
discrepant by $>3\sigma$ with the observed signal. Therefore the
observed signal \emph{cannot} be due to a selection effect. We concur
with Refs.~\cite{Hunt:2008wp,Inoue:2010rp} that, if the observed signal 
is due to the ISW effect, deep superstructures appear to be far
more numerous than expected in a $\Lambda$CDM cosmology.

This differs from the conclusion of Ref.~\cite{Papai:2010gd} who also used
similar density profiles. We have demonstrated that this difference 
is not due to the small difference in profiles used, and have argued that 
it is instead because those authors interpreted at face value their 
template-fitting method when applied for density contrasts where the 
template necessarily breaks down for voids (i.e., requires $\delta<-1$!). 
A more appropriate interpretation is entirely compatible with the results 
presented here.

An interesting question is whether the expected signal of the most
extreme superstructures in a $\Lambda$CDM universe is possible to
detect in principle. For $50$ $4^\circ$ circles of the CMB, 
Ref.~\cite{Granett:2008ju} reports the statistical uncertainty is $3.1\;\mu$K. 
Hence the expected statistical uncertainty in the average for a sample of 
$N$ such patches will be approximately $(3.1\times\sqrt{50}/\sqrt{N})\;\mu$K, 
so that for $N\simeq3000$, the uncertainty on a measurement of 
$\langle\Delta T\rangle$ is $\Delta T_\mathrm{noise}\simeq0.4\;\mu$K. 
Thus a detection of $|\langle\Delta T\rangle|=1\;\mu$K can be made with 
roughly $2.5\sigma$ significance, averaged over an ensemble of 3000 
superstructures. From Fig.~\ref{fig:Tlargest} it is seen that for the 3000 largest
voids in the SDSS DR6 survey volume, $|\langle\Delta
T\rangle|\sim1\;\mu$K which is of the right order of magnitude but
somewhat too small for detection at high significance. However the
SDSS window is not large enough to contain 3000 independent $4^\circ$
patches on the sky so in any case a larger survey would be needed in
order to measure a statistically significant signal and this would
contain more supervoids. This order-of-magnitude estimate indicates
that even if the reported observation is a statistical anomaly or flawed in some
other respect, the ISW imprint of superstructures in a $\Lambda$CDM 
cosmology may be large enough to be detected in future surveys.

As noted earlier, the detection of the ISW imprint of individual
superstructures provides an important complement to full-sky
CMB-galaxy cross-correlation studies. It has the potential to provide
information about the radii, density contrasts and density profiles of
specific structures in the dark matter distribution that lie in the extreme 
tail of the probability distribution function. Our calculation demonstrates 
that the predicted ISW signal from the most extreme superstructures is 
far too small to explain the temperature fluctuations seen in 
Ref.~\cite{Granett:2008ju}, and that therefore the reported observations 
are in tension with the $\Lambda$CDM model. Further observational 
investigation of this issue is warranted.

Nevertheless, the standard cosmology successfully fits many other 
observations, and any resolution of the discrepancy discussed here 
must also satisfy these constraints. A possible physical explanation 
would be that the primordial perturbations are non-gaussian. This 
would influence both the abundance of these extreme regions (e.g., 
\cite{Matarrese:2000iz,Kamionkowski:2008sr}) as well as their density 
profile, thus changing their expected ISW signal. A primordial non-gaussianity 
disproportionately affects the tails of the probability distribution function. 
Since the observation in Ref.~\cite{Granett:2008ju} specifically probes 
the ISW effect of structures in these tails, it is expected to be more sensitive 
to non-gaussianity than the standard cross-correlation ISW studies. Interestingly, 
several cross-correlation studies also find an enhancement of 
the ISW signal over $\Lambda$CDM expectations, e.g. 
\cite{Ho:2008,Giannantonio:2008zi,Goto:2012yc}, though given the 
large error bars this discrepancy has not been considered too significant. 
Therefore, primordial non-gaussianity may be able to explain this signal 
while preserving the success of $\Lambda$CDM on other fronts. However, 
this would then undermine the use of the ISW effect as an independent test 
for $\Lambda$.

It is worth noting that, qualitatively, a similar enhancement is found for 
overdensities as well as for voids, though we did not quantify this. As 
pointed in Ref. \cite{Enqvist:2010bg} a primordial 
skewness, parameterised by $f_\mathrm{NL}$, would not be able to 
enhance the abundance of both over- and underdense regions 
simultaneously; however a primordial kurtosis, parameterised by a 
positive $g_\mathrm{NL}$, would indeed do so, and be less constrained by 
the CMB. 

Another possible explanation might lie in a modification of the growth rate of 
perturbations as can happen in, e.g., models based on scalar-tensor 
gravity \cite{Nagata:2003qn}. Such models may also change predictions by
altering the behaviour of photons in gravitational potentials. The presence 
of large-scale inhomogeneities can themselves alter the growth rate and 
this too deserves further attention.

\section{Acknowledgements}
We would like to thank the anonymous referee for highlighting several 
areas where our arguments were unclear and for suggesting 
improvements. We thank Syksy R\"as\"anen, Ben Hoyle and Tom Shanks 
for helpful comments and discussion, and Phil Bull for assistance with 
producing the figures. SH is supported by the Academy of Finland grant 131454.

\end{document}